\documentclass[12pt]{article}
\newlength{\extraspace}
\newlength{\extraspaces}
\newcommand{\be}{\begin{equation}
\addtolength{\abovedisplayskip}{\extraspaces}
\addtolength{\belowdisplayskip}{\extraspaces}
\addtolength{\abovedisplayshortskip}{\extraspace}
\addtolength{\belowdisplayshortskip}{\extraspace}}
\newcommand{\ee}{\end{equation}}

\newcommand{\ba}{\begin{eqnarray}
\addtolength{\abovedisplayskip}{\extraspaces}
\addtolength{\belowdisplayskip}{\extraspaces}
\addtolength{\abovedisplayshortskip}{\extraspace}
\addtolength{\belowdisplayshortskip}{\extraspace}}
\newcommand{\ea}{\end{eqnarray}}

\newcounter{saveeqn}

\newcommand{\dif}{\mathrm{d}}
\newcommand{\me}{\mathrm{e}}

\begin{document}
\addtolength{\baselineskip}{1.5mm}

\thispagestyle{empty}
\begin{flushright}
hep-th/0412164\\
\end{flushright}
\vbox{}
\vspace{2.5cm}

\begin{center}
{\LARGE{A no-go theorem for accelerating cosmologies\\[2mm]
from M-theory compactifications
        }}\\[16mm]
{Edward Teo}
\\[6mm]
{\it Department of Physics,
National University of Singapore, 
Singapore 119260}\\[15mm]

\end{center}
\vspace{2cm}

\centerline{\bf Abstract}
\bigskip 
\noindent 
It is known that four-dimensional cosmologies exhibiting transient 
phases of acceleration can be obtained by compactifications of low-energy 
effective string or M-theory on time-varying manifolds. In the
four-dimensional theory, the acceleration can be attributed to
a quintessential scalar field with a positive effective potential. 
Recently, Townsend has conjectured that the potentials obtained by such 
compactifications cannot give rise to late-time accelerating 
universes which possess future event horizons. Such a `no-go' result 
would be desirable, since current string or M-theory seems unable to 
provide an adequate description of space-times with future event horizons. 
In this letter, we provide a proof of this conjecture for a class of 
warped compactifications with a single scalar modulus parametrising 
the volume of the compactification manifold.


\newpage

Recent observations indicate that the universe is undergoing an 
accelerated expansion, but the nature and origin of the `dark energy' 
driving this acceleration remains elusive. One of the more popular 
candidates is a scalar field with a positive potential, also known 
as quintessence. While it is easy to construct ad hoc potentials
which give either transient or eternal acceleration, it remains
a problem to obtain suitable potentials from a more fundamental 
theory, such as string or M-theory. Part of the problem stems from
the existence of a `no-go' theorem \cite{Gibbons} for (possibly warped) 
compactifications of 10- or 11-dimensional string or M-theory 
on a time-independent manifold. This no-go theorem states that if 
the strong energy condition (SEC) is satisfied in the higher-dimensional 
theory, as it is in low-energy effective string or M-theory, 
then it is also satisfied in
the compactified theory. The latter fact will rule out the 
possibility of accelerating universes in the compactified theory,
since acceleration requires a violation of the SEC.

A way to circumvent the no-go theorem was found by Townsend
and Wohlfarth \cite{TW}, who considered hyperbolic
compactification manifolds that have a {\it time-varying\/} 
volume. They showed that the resulting four-dimensional universe
will exhibit a transient phase of acceleration. This result was 
subsequently extended by others to include product space
compactifications \cite{Chen}, 
as well as flux compactifications, i.e., compactifications
of higher-dimensional antisymmetric tensor fields with non-zero flux
\cite{Ohta} (see also \cite{Costa} for some earlier work, and 
\cite{Ohta1} for a recent review).
Thus, it appears that accelerating universes are quite generic 
in time-dependent compactifications of higher-dimensional supergravity 
theories, which usually contain such antisymmetric tensor fields.

In the reduced four-dimensional space-time, the volume modulus of the
compactification manifold appears as a scalar field $\phi$, with 
a potential $V(\phi)$ whose exact form depends on the details of the
compactification. For acceleration to occur, this potential has to
be positive, or at least positive in some region of the space of scalar 
fields. An example of a potential which commonly arises is a single 
exponential of the form
\be
\label{exponential}
V=\Lambda\,\me^{-a\phi},
\ee
where $\Lambda$ and $a$ are positive constants. It is known that 
compactifications on hyperbolic manifolds of constant curvature
yield exponential potentials with $1<a<\sqrt{3}$, while flux
compactifications yield exponential potentials with 
$\sqrt{3}\leq a<3$. However, no conventional supergravity 
compactifications are known to give exponential potentials with 
$a\leq1$.\footnote{More exotic types of compactifications, such 
as when the compactification manifold is non-compact, could give rise
to such exponential potentials \cite{Townsend1}. Another way is
by gauging a scaling symmetry of the metric, as was
very recently considered in \cite{Brax}. But these cases
are not covered by the original no-go theorem \cite{Gibbons},
nor the one presented in this letter.}

It has been shown in \cite{Hellerman} that exponential 
potentials (\ref{exponential}) with $a<1$ will give rise to 
eternally accelerating cosmologies with future event
horizons. The presence of future event horizons means that current 
string or M-theory may not be able to provide an adequate
description of such cosmologies, in the same way that it is unable to 
provide a description of de Sitter space (see, e.g., \cite{Banks}).
This has prompted Townsend \cite{Townsend} to make the conjecture that 
universes with future event horizons cannot arise from compactifications 
of low-energy effective string or M-theory. In particular, this will
rule out exponential potentials with $a<1$. In this letter, we will
provide a proof of this conjecture, which may be regarded as a
type of no-go theorem that generalises the original one 
of \cite{Gibbons} to time-dependent compactifications.

Our starting point is a $D$-dimensional theory satisfying the
SEC. The specific examples we have 
in mind are the $D=10$ or 11 supergravity actions that serve as the 
low-energy effective actions for string or M-theory. Since our 
intention is to compactify this theory to four dimensions, we
consider a space-time metric of the form:
\be
\label{D_metric}
\dif s^2_{D}=\Omega^2(y)\,\me^{-\sqrt{\frac{n}{n+2}}\phi(x)}\dif s^2_{4}(x)
+\me^{\frac{2}{\sqrt{n(n+2)}}\phi(x)}\dif s^2_{n}(y)\,,
\ee
where $n=D-4$. Here, $\dif s^2_{4}$ is the metric of the four-dimensional 
space-time, while $\dif s^2_{n}$ is the metric of some compact, non-singular 
$n$-manifold ${\cal M}$ without boundary.
$\Omega$ is a smooth, non-vanishing field on ${\cal M}$, known as
a warp factor. On the other hand, $\phi$ is
a four-dimensional scalar field which parametrises the volume of ${\cal M}$. 
The exponents in (\ref{D_metric}) have been chosen to give a 
four-dimensional metric that is in the Einstein conformal frame, as well 
as a canonically normalised kinetic term for $\phi$ in the four-dimensional 
effective action. In the case when $\phi$ is identically zero, we recover 
the ans\"atz used in \cite{Gibbons}.

As we are specifically interested in four-dimensional cosmological 
solutions arising from time-dependent compactifications, let us write 
the four-dimensional metric in the standard FRW form:
\be
\label{4_metric}
\dif s^2_{4}=-\dif\tau^2+S^2(\tau)\left(\frac{\dif r^2}{1-kr^2}
+r^2\dif\Omega^2_2\right),
\ee
where $S(\tau)$ is a time-dependent scale factor, and $k=+1,0,-1$,
corresponding to closed, flat and open universes, respectively.
Also, the scalar field $\phi=\phi(\tau)$ is assumed to be a function 
of time only.

The four-dimensional action resulting from the compactification
of the $D$-dimensional theory will take the form
(possibly after consistent truncations of irrelevant fields): 
\be
\label{4_action}
\frac{1}{16\pi G_4}\int\dif^4x\sqrt{-g}\left[R
-\frac{1}{2}\partial_\mu\phi\partial^\mu\phi-V(\phi)\right],
\ee
for some effective scalar potential $V(\phi)$. The exact form of this 
potential depends on the details of the compactification; examples
that are known to arise from string or M-theory include 
exponential ones of the form (\ref{exponential}) as discussed
above, and the sum of two exponential terms as studied in 
\cite{Ohta,Emparan,Neupane,Jarv}.
The equations of motion that follow from the action (\ref{4_action}), after
substituting the form of the metric (\ref{4_metric}), are
\ba
\label{S_eqn}
\frac{\ddot S}{S}&=&\frac{1}{6}(-\dot\phi^2+V)\,,\\
\label{phi_eqn}
\ddot\phi&=&-3\dot\phi\frac{\dot S}{S}-V',
\ea
where $\dot{}\equiv\frac{\dif}{\dif\tau}$ and 
${}'\equiv\frac{\dif}{\dif\phi}$. We also have a third equation of 
motion, the so-called Friedmann equation, but it is not needed here 
[c.f.\ (\ref{Friedmann}) below].

Now, the SEC on the matter content of the $D$-dimensional theory
implies that the $D$-dimensional Ricci tensor has a time-time
component which satisfies $R^{(D)}_{00}\geq0$. A calculation using 
the metrics (\ref{D_metric}) and (\ref{4_metric}) yields the explicit 
expression:
\be
\label{R00}
R^{(D)}_{00}=-3\frac{\ddot S}{S}-\frac{1}{2}\dot\phi^2
+\frac{1}{2}\sqrt{\frac{n}{n+2}}\left(3\dot\phi\frac{\dot S}{S}
+\ddot\phi\right)+\frac{1}{4}\me^{-\sqrt{\frac{n+2}{n}}\phi}\,
\Omega^{-2}(y)\nabla_y^2\Omega^4(y)\,.
\ee
Multiplying this by $\Omega^2$ and integrating over the higher-dimensional
$n$-manifold ${\cal M}$, we obtain
\ba
\int_{\cal M}\Omega^2R^{(D)}_{00}&=&\left[\int_{\cal M}\Omega^2\right]
\left[-3\frac{\ddot S}{S}-\frac{1}{2}\dot\phi^2
+\frac{1}{2}\sqrt{\frac{n}{n+2}}\left(3\dot\phi\frac{\dot S}{S}
+\ddot\phi\right)\right]\cr
&=&\left[\int_{\cal M}\Omega^2\right]
\left(-\frac{1}{2}V-\frac{1}{2}\sqrt{\frac{n}{n+2}}\,V'\right),
\ea
where we have used the two equations of motion (\ref{S_eqn}) and 
(\ref{phi_eqn}) to rewrite the right-hand side in the second line.
The $D$-dimensional SEC then becomes a very simple condition on the 
potential $V$:
\be
\label{condition}
V'\leq-\sqrt{\frac{n+2}{n}}\,V\,.
\ee
Although we are primarily interested in string or
M-theory (for which $n=6$ or 7), we shall keep $n$ arbitrary. 
For the case of an exponential potential
(\ref{exponential}), this translates to the condition that $a>1$. 
This immediately explains the observation made in \cite{Townsend}
that exponential potentials with $a<1$ do not seem to arise from 
compactifications of string or M-theory. If such potentials could 
arise, they would imply a violation of the SEC in these 
higher-dimensional theories.

Our task now is to show that the condition (\ref{condition}) 
will ensure that a four-dimensional universe which solves the
equations of motion coming from (\ref{4_action}),
will not possess any future event horizons even if it undergoes 
late-time acceleration. We start off by recalling what is 
known about the general behaviour of $V$. It has been argued in 
\cite{Giddings} that $V$ must generically vanish as $\phi$ 
approaches infinity, but that it could still have local minima; 
a few possible shapes for $V$ are illustrated in \cite{Giddings}. 
In particular, a positive 
minimum would imply the presence of a metastable de Sitter region,
while a negative minimum would correspond to an anti-de Sitter
`basin of attraction'. The latter would cause the universe to
evolve into a big crunch singularity, and so can be excluded
from our consideration without any loss of generality.

It can been seen that the condition (\ref{condition}) immediately rules out
the possibility of any potential arising from compactification having 
stationary points with $V>0$.\footnote{This point was also noted in 
\cite{Townsend}, and references therein.} 
Potentials having stationary points with $V<0$ are not ruled out by 
this condition---indeed they are present in flux compactifications
on a sphere (see, e.g., \cite{Emparan})---but we shall not  
consider such potentials any further as they will have one or more 
AdS basins of attraction. 
Finally, potentials having stationary points with $V=0$ are also 
allowed by (\ref{condition}), but they are necessarily either local 
maxima or saddle points. Both cases will lead to AdS basins of attraction,
and so again we shall exclude such potentials from consideration. 
Hence, we are left with strictly positive potentials that have no 
stationary points.

Since $V$ is assumed to be positive from now on, the condition
(\ref{condition}) can be recast as
\be
\label{condition1}
\frac{V'}{V}<-1\,.
\ee
Note that this implies that $V$ is falling to zero {\it faster\/} 
than the $a=1$ exponential potential [which saturates the bound in
(\ref{condition1})]. It is now possible to see, at least intuitively,
why Townsend's conjecture is true, if we recall that
cosmological solutions can be interpreted as a ball rolling, with 
friction, up and down the potential $V$ \cite{Emparan}. 
A steeper descent will imply a late-time evolution in which the 
scalar field generally has a larger kinetic energy as compared to its 
potential energy, thus leading to a universe with a lower rate of 
acceleration by (\ref{S_eqn}). Since it is known that the $a=1$ 
exponential potential gives rise to universes that are free of
future event horizons \cite{Hellerman,Boya}, it follows that 
any positive potential $V$ satisfying (\ref{condition1}) will also 
give rise to universes without future event horizons.\footnote{The 
$a=1$ exponential potential was also used as a `reference potential'
in \cite{TW1} to study the asymptotic behaviour of cosmologies with 
more general potentials.} 

In the remainder of this letter, we shall see in detail how the preceding
result follows from a phase-space analysis of the equations of motion.
This approach has the added bonus in that it will allow us to 
classify the various asymptotic
behaviour that can arise from different $V$. Let us define a new time 
coordinate $t$, related to the proper time $\tau$ by
\be
\label{t}
\dif t=\sqrt{V}\,\dif\tau\,.
\ee
If we further write $S=\me^{\alpha}$, then the equations of 
motion (\ref{S_eqn}) and (\ref{phi_eqn}) become 
\ba
\label{eom1}
\ddot\alpha&=&\frac{1}{6}(1-\dot\phi^2)-\dot\alpha^2
-\frac{V'}{2V}\dot\alpha\dot\phi\,,\\
\label{eom2}
\ddot\phi&=&-3\dot\alpha\dot\phi-\frac{V'}{2V}(2+\dot\phi^2)\,,
\ea
where we denote $\dot{}\equiv\frac{\dif}{\dif t}$ from now on.
Note that the definition (\ref{t}) has deliberately been chosen
to ensure that the potential appears in (\ref{eom1})
and (\ref{eom2}) only as the ratio $\frac{V'}{V}$.
These two differential equations govern the time evolution of the
solution, and the result can be plotted as a trajectory in the
$(\dot\phi,\dot\alpha)$-phase space.
We also have the Friedmann equation:
\be
\label{Friedmann}
\dot\alpha^2-\frac{1}{12}\dot\phi^2=\frac{1}{6}-\frac{k\,\me^{-2\alpha}}
{V}\,.
\ee
When $k=0$, (\ref{Friedmann}) describes a hyperbola in the
phase space, which divides the latter
into regions where $k=\pm1$. We would mainly be interested in 
the upper half of the phase space in which $\dot\alpha>0$, 
corresponding to expanding universes. Furthermore, since the
second derivative of the scale factor $S$ with respect to
proper time $\tau$ is given by
\be
\frac{\dif^2S}{\dif\tau^2}=\frac{VS}{6}(1-\dot\phi^2)\,,
\ee 
we see that positive acceleration occurs when $|\dot\phi|<1$, 
corresponding to a vertical open strip in the phase space.

\begin{table}
\begin{center}
\begin{tabular}{cccl}
\hline
\hline
\noalign{\smallskip}
$k$ & $a$ & fixed point & late-time behaviour\\
\noalign{\smallskip}
\hline
\noalign{\smallskip}
closed ($k=+1$) & $a<1$ &1& accel.\ with future horizon\\
\noalign{\smallskip}
            && 2&accel. or decel.\\
\noalign{\smallskip}
\hline
\noalign{\smallskip}
flat ($k=0$) & $a<1$ &1& accel.\ with future horizon\\
\noalign{\smallskip}
            & $a=1$ &1=2& accel. or decel.\\
\noalign{\smallskip}
            & $1<a<\sqrt{3}$ &1& decel.\\
\noalign{\smallskip}
\hline
\noalign{\smallskip}
open ($k=-1$) & $a<1$ &1& accel.\ with future horizon\\
\noalign{\smallskip}
            & $a=1$ &1=2& accel.\\
\noalign{\smallskip}
            & $1<a\leq\frac{2}{\sqrt{3}}$ &2& accel.\ or decel.\\
\noalign{\smallskip}
            & $a>\frac{2}{\sqrt{3}}$ &2& oscillating accel.\ and decel.\\
\noalign{\smallskip}
\hline
\hline
\end{tabular}
\caption{Possible late-time behaviour of expanding universes with an 
exponential scalar potential (\ref{exponential}) that are attracted 
to the fixed points (\ref{fp1}) or (\ref{fp2}).}
\end{center}
\end{table}

The dynamical system (\ref{eom1}) and (\ref{eom2})
is particularly simple for the case of an exponential potential 
(\ref{exponential}), and has been well-studied by various authors 
(see, e.g., \cite{Halliwell,Vieira,Jarv}). In this case, 
there are two fixed points in the upper-half phase space, given by
\ba
\label{fp1}
(\dot\phi_1,\dot\alpha_1)&=&\left(\frac{\sqrt{2}\,a}{\sqrt{3-a^2}},\frac{1}{\sqrt{2(3-a^2)}}\right),\\
\label{fp2}
(\dot\phi_2,\dot\alpha_2)&=&\left(1,\frac{a}{2}\right).
\ea
Note that the first lies on the $k=0$ hyperbola, while the 
second lies on the boundary between the acceleration and deceleration
regions. They coincide 
when $a=1$. Table 1 summarises the late-time behaviour of all
solutions that are attracted to either of these fixed points, 
specifically whether they undergo accelerated and/or decelerated 
expansion. Whenever a universe undergoing accelerated expansion
possesses a future event horizon, this is indicated in the table.
As can be seen, such horizons are only present in the case when 
$a<1$ \cite{Hellerman,Boya}, corresponding to the situation where the
fixed point lies on the part of the hyperbola {\it inside\/} the 
acceleration region. Other solutions that are not attracted
to either of these fixed points need not be considered, as they
correspond to universes which will either undergo late-time 
decelerated expansion or contract into a big crunch singularity
\cite{Jarv}.

We return to the case of a general positive potential $V$
satisfying (\ref{condition1}). Since it
has no stationary points, $\phi$ will exhibit a runaway 
behaviour $\phi\rightarrow\infty$ at late times. Suppose the 
dynamical equations 
(\ref{eom1}) and (\ref{eom2}) have fixed points\footnote{Called 
quasi fixed points in the terminology of \cite{Jarv}.} in this limit.
This means that the ratio $\frac{V'}{V}$ will approach
some constant value:
\be
\lim_{\phi\rightarrow\infty}\frac{V'(\phi)}{V(\phi)}\equiv-a_\infty\,,
\ee
which satisfies $a_\infty\geq1$. Thus, solutions
that are attracted to these fixed points have the {\it same\/} asymptotic
behaviour as those for the exponential potential (\ref{exponential}) 
with $a=a_\infty$. In particular, the fixed points (\ref{fp1}) and (\ref{fp2}) 
will continue to serve as attractors for expanding universes with 
such potentials, and the late-time behaviour of 
these solutions coincide with those listed in Table 1 for $a\geq1$. It 
follows that those universes which are undergoing late-time acceleration 
are free of future event horizons. Again, we need not consider solutions 
that are not attracted to these fixed points, as they correspond to
universes which will either undergo late-time 
decelerated expansion or contract into a big crunch singularity.

We now turn to the possibility that (\ref{eom1}) and (\ref{eom2}) 
do not admit any fixed points; this would occur when 
$\frac{V'}{V}$ does not have a well-defined asymptotic limit (e.g.,
it is asymptotically periodic) or when it diverges to $-\infty$.
A phase-space trajectory in this case will therefore not terminate. 
However, its general direction of flow can still be inferred from those 
of the exponential potential with exponents $a>1$, since the former
can be viewed as arising from an exponential potential with an 
effective exponent $a(\phi)$ that varies along the trajectory 
according to the value of $-\frac{V'}{V}$ [which is 
always greater than 1 by (\ref{condition1})]. With this fact in
mind, we consider the three possibilities $k=0,\pm1$ separately.

The situation is simplest for closed ($k=+1$) universes, for which 
it is known that any trajectory with
exponential potential $a>1$ will flow to the lower-half of the phase
space (c.f.\ the phase portraits in \cite{Halliwell,Vieira,Jarv}), 
corresponding to a universe that 
is contracting into a big crunch singularity. It therefore follows 
that any solution with positive potential satisfying (\ref{condition1}) 
will also do likewise, and therefore can be excluded from consideration.

For flat ($k=0$) universes, any trajectory is restricted to flow along
the (upper) hyperbola described below (\ref{Friedmann}). If 
$a(\phi)\geq\sqrt{3}$
asymptotically, then the trajectory will flow to right infinity, 
corresponding to an expanding but decelerating universe. Otherwise, 
the trajectory would asymptotically oscillate back and forth along the 
right side of the hyperbola. In any case, it cannot enter the 
acceleration region $|\dot\phi|<1$ from the right, since it follows from 
(\ref{condition1}) and (\ref{eom2}) that $\ddot\phi>0$ at $\dot\phi=1$. 
Thus, these trajectories will also describe expanding but decelerating 
universes, which would not possess any future event horizons.

The most interesting situation occurs for open ($k=-1$) universes. If
$a(\phi)\rightarrow\infty$ asymptotically, then it can be seen from
the phase portraits in \cite{Halliwell,Vieira,Jarv} that
the trajectory will flow to the top of the phase space, approaching 
$\dot\phi=1$ from the right. This corresponds to an expanding but 
decelerating universe. Otherwise, the trajectory will asymptotically
oscillate between the acceleration and deceleration regions in the
top-right quadrant of the phase space, corresponding to a 
universe undergoing late-time oscillating acceleration and deceleration.
To show that such a universe would not possess a future event
horizon, it suffices to show that the integral $\int_{\tau_0}^\infty
\frac{\dif\tau}{S(\tau)}$ diverges (see, e.g., the second reference
in \cite{Hellerman} or \cite{Carroll}).
Indeed, it follows from (\ref{t}) and (\ref{Friedmann}) that
\be
\int_{\tau_0}^\infty\frac{\dif\tau}{S}=
\int_{\phi_0}^\infty\frac{\dif\phi}{\left(\frac{\dif\phi}{\dif\tau}\right)S}
=\int_{\phi_0}^\infty\frac{1}{\dot\phi}
\sqrt{\dot\alpha^2-\hbox{$\frac{1}{12}$}\dot\phi^2
-\hbox{$\frac{1}{6}$}}\,\dif\phi\,,
\ee
which is clearly divergent since the term under the square-root does
not vanish asymptotically. 
Hence, none of these universes will possess a future event horizon, 
thereby completing the proof of Townsend's conjecture.

Now, since the condition (\ref{condition1}) is valid for all times,
it can tell us more than just the late-time behaviour of 
the acceleration of the universe. For example, it can be used
to constrain the amount of cosmological inflation that is possible
in the early universe. Indeed, (\ref{condition1}) immediately implies that
the `slow-roll' condition $|\frac{V'}{V}|\ll1$ cannot be satisfied
for such models. In other words, the potential $V$ is falling to
zero too quickly and any inflationary era will be too short to be
realistic. This is consistent with observations made previously in 
\cite{Chen,Wohlfarth1} that the number of e-foldings that can be obtained 
from such models is only of order 1. However, this is still adequate
to describe the current phase of acceleration that our universe
seems to be undergoing.

In this letter, we have only considered compactifications
of low-energy effective string or M-theory. Although this has been
the focus of much of the research in string cosmology, there has been
some recent progress to include string corrections, both perturbative
and non-perturbative, as well as the presence of extended sources such 
as branes. This would certainly modify the form of $V$. 
In some cases, the resulting potentials have been shown to possess a 
metastable de Sitter region \cite{Kachru}. The no-go theorem
presented in this letter is not surprisingly violated for such examples, 
and it would be interesting to see if a more general one exists
that would encompass these cases.

\bigbreak\bigskip\bigskip\centerline{{\bf Acknowledgement}}
\nobreak\noindent I would like to thank Paul Townsend for some 
helpful comments.

\bigskip\bigskip

{\renewcommand{\Large}{\normalsize}
}

\begin{thebibliography}{99}

\bibitem{Gibbons}
G.~W.~Gibbons, 
in {\it Supersymmetry, Supergravity and Related Topics\/}, 
eds. F. de Aguila, J.~A.~de Azc\'arraga and L.~E.~Iba\~nez,
(World Scientific, Singapore, 1985);
J.~M.~Maldacena and C.~Nu\~nez,
Int.\ J.\ Mod.\ Phys.\ A {\bf 16} (2001) 822
[arXiv:hep-th/0007018].

\bibitem{TW}
P.~K.~Townsend and M.~N.~R.~Wohlfarth,
Phys.\ Rev.\ Lett.\  {\bf 91} (2003) 061302
[arXiv:hep-th/0303097].

\bibitem{Chen}
C.~M.~Chen, P.~M.~Ho, I.~P.~Neupane and J.~E.~Wang,
JHEP {\bf 0307} (2003) 017
[arXiv:hep-th/0304177];
C.~M.~Chen, P.~M.~Ho, I.~P.~Neupane, N.~Ohta and J.~E.~Wang,
JHEP {\bf 0310} (2003) 058
[arXiv:hep-th/0306291].

\bibitem{Ohta}
N.~Ohta,
Phys.\ Rev.\ Lett.\  {\bf 91} (2003) 061303
[arXiv:hep-th/0303238];
S.~Roy,
Phys.\ Lett.\ B {\bf 567} (2003) 322
[arXiv:hep-th/0304084];
M.~N.~R.~Wohlfarth,
Phys.\ Lett.\ B {\bf 563} (2003) 1
[arXiv:hep-th/0304089];
N.~Ohta,
Prog.\ Theor.\ Phys.\  {\bf 110} (2003) 269
[arXiv:hep-th/0304172].

\bibitem{Costa}
L.~Cornalba and M.~S.~Costa,
Phys.\ Rev.\ D {\bf 66} (2002) 066001
[arXiv:hep-th/0203031].

\bibitem{Ohta1}
N.~Ohta,
``Accelerating cosmologies and inflation from M/superstring theories,''
arXiv:hep-th/0411230.

\bibitem{Townsend1}
P.~K.~Townsend,
JHEP {\bf 0111} (2001) 042
[arXiv:hep-th/0110072].

\bibitem{Brax}
Ph.~Brax, C.~van de Bruck and A.~C.~Davis,
``Cosmic acceleration in massive half-maximal supergravity,''
arXiv:hep-th/0411208.

\bibitem{Hellerman}
S.~Hellerman, N.~Kaloper and L.~Susskind,
JHEP {\bf 0106} (2001) 003
[arXiv:hep-th/0104180];
W.~Fischler, A.~Kashani-Poor, R.~McNees and S.~Paban,
JHEP {\bf 0107} (2001) 003
[arXiv:hep-th/0104181].

\bibitem{Banks}
T.~Banks,
``Cosmological breaking of supersymmetry or little Lambda goes back to  the
future. II,''
arXiv:hep-th/0007146.

\bibitem{Townsend}
P.~K.~Townsend,
``Cosmic acceleration and M-theory,''
arXiv:hep-th/0308149.

\bibitem{Emparan}
R.~Emparan and J.~Garriga,
JHEP {\bf 0305} (2003) 028
[arXiv:hep-th/0304124].

\bibitem{Neupane}
I.~P.~Neupane,
Class.\ Quant.\ Grav.\  {\bf 21} (2004) 4383
[arXiv:hep-th/0311071].

\bibitem{Jarv}
L.~J\"arv, T.~Mohaupt and F.~Saueressig,
JCAP {\bf 0408} (2004) 016
[arXiv:hep-th/0403063].

\bibitem{Giddings}
S.~B.~Giddings,
Phys.\ Rev.\ D {\bf 68} (2003) 026006
[arXiv:hep-th/0303031];
S.~B.~Giddings and R.~C.~Myers,
Phys.\ Rev.\ D {\bf 70} (2004) 046005
[arXiv:hep-th/0404220].

\bibitem{Boya}
L.~J.~Boya, M.~A.~Per and A.~J.~Segui,
Phys.\ Rev.\ D {\bf 66} (2002) 064009
[arXiv:gr-qc/0203074].

\bibitem{TW1}
P.~K.~Townsend and M.~N.~R.~Wohlfarth,
Class.\ Quant.\ Grav.\  {\bf 21} (2004) 5375
[arXiv:hep-th/0404241].

\bibitem{Halliwell}
J.~J.~Halliwell,
Phys.\ Lett.\ B {\bf 185} (1987) 341.

\bibitem{Vieira}
P.~G.~Vieira,
Class.\ Quant.\ Grav.\  {\bf 21} (2004) 2421
[arXiv:hep-th/0311173].

\bibitem{Carroll}
M.~Trodden and S.~M.~Carroll,
``TASI lectures: Introduction to cosmology,''
arXiv:astro-ph/0401547.

\bibitem{Wohlfarth1}
M.~N.~R.~Wohlfarth,
Phys.\ Rev.\ D {\bf 69} (2004) 066002
[arXiv:hep-th/0307179].

\bibitem{Kachru}
S.~Kachru, R.~Kallosh, A.~Linde and S.~P.~Trivedi,
Phys.\ Rev.\ D {\bf 68} (2003) 046005
[arXiv:hep-th/0301240];
S.~Kachru, R.~Kallosh, A.~Linde, J.~Maldacena, L.~McAllister and S.~P.~Trivedi,
JCAP {\bf 0310} (2003) 013
[arXiv:hep-th/0308055].


\end{thebibliography}
\end{document}